# From the evolution of public data ecosystems to the evolving horizons of the forward-looking intelligent public data ecosystem empowered by emerging technologies

Anastasija Nikiforova[1], Martin Lnenicka[2], Petar Milić[3], Mariusz Luterek4 and Manuel Pedro Rodríguez Bolívar[5]

[1] University of Tartu, Estonia
[2] University of Pardubice, Pardubice, Czech Republic
[3] University of Pristina – Kosovska Mitrovica, Kosovska Mitrovica, Serbia
[4] University of Warsaw, Warsaw, Poland
[5] University of Granada, Granada, Spain

Corresponding author: Anastasija Nikiforova, Anastasija.Nikiforova@ut.ee

**Abstract.** *Public data ecosystems (PDEs) represent complex socio-technical systems crucial for optimizing data use in the public sector and outside it. Recognizing their multifaceted nature, previous research proposed a six-generation Evolutionary Model of Public Data Ecosystems (EMPDE). Designed as a result of a systematic literature review on the topic spanning three decade, this model, while theoretically robust, necessitates empirical validation to enhance its practical applicability. This study addresses this gap by validating the theoretical model through a real-life examination in five European countries - Latvia, Serbia, Czech Republic, Spain, and Poland. This empirical validation provides insights into PDEs dynamics and variations of implementations across contexts, particularly focusing on the 6th generation of forward-looking PDE generation named "Intelligent Public Data Generation" that represents a paradigm shift driven by emerging technologies such as cloud computing, Artificial Intelligence, Natural Language Processing tools, Generative AI, and Large Language Models (LLM) with potential to contribute to both automation and augmentation of business processes within these ecosystems. By transcending their traditional status as a mere component, evolving into both an actor and a stakeholder simultaneously, these technologies catalyze innovation and progress, enhancing PDE management strategies to align with societal, regulatory, and technical imperatives in the digital era.*

**Keywords:** *AI, Artificial Intelligence, Emerging Technologies, Generative AI, Open Government Data, Open Data Ecosystem, Public Data Ecosystem, Socio-technical systems, Systems theory, Evolution Model.*

## 1 Introduction

The public sector confronts multifaceted societal, regulatory, and technical challenges due to the burgeoning volumes of available data and the imperative to optimize their utilization for enhanced efficiency and effectiveness [1-2]. Understanding the data-related elements that are affected and contribute to the system development is paramount to ensuring alignment with intended purposes [3]. Over the past decade, the ecosystem perspective has gained prominence, emphasizing the interconnectedness of social, technological, and information systems, and their self-organizing and co-evolutionary nature, particularly pertinent in understanding public data ecosystems (PDEs) [4-10]. Similar to digital or Information and communication technology (ICT) ecosystems, PDE heavily rely on ICT and digital resources with data as the primary asset, facilitating data flows, technological deployment, stakeholder interactions, and management [11-16]. These ecosystems vary in objectives, composition of elements from which they are composed, environments in which they are built, and influential factors, necessitating tailored approaches [3,6,10,17]. As such, PDE can be defined as *a public data ecosystem can be defined as a dynamic network comprising interconnected elements that enable a range of data-related activities that encompass the entire data lifecycle, from collection and management to sharing and reuse, involving diverse stakeholders with varied objectives*.

Spatial Data Infrastructures (SDIs) and open (government) data ecosystems have garnered significant attention, particularly in facilitating data exchange, sharing, and accessibility, supported by geodata portals and open data portals [7,16,18,20-21]. These ecosystems, their constituent elements, and relationships between them are dynamic and evolve with time [14-15] in response to technological advancements, societal needs, and regulatory frameworks [13,17,20], requiring ongoing assessment and adaptation [13-15,17,20]. This evolution can be explored across different administrative levels, including the global context, and is characterized by elements affecting and driving it that may vary in importance and impact over time [13,17,20].

Understanding the dynamics and evolution of public data ecosystems, including their formative characteristics and relationships between them is vital for their stability and the success of data-driven projects [22]. This, as shown by Heimstädt et al. [12], allows us to trace the ecosystem's development step-by-step and pinpoint key turning points, which may be both internal and external factors, e.g., technological developments such as cloud computing, the Internet of Things (IoT), big data and Artificial Intelligence (AI) that drive the digital transformation at different levels [23].

To comprehensively explore and characterize the interrelated and interdependent components and relationships between them within PDE, a systems theory approach is invaluable. This approach facilitates the understanding of the context, structure, function, role, and behaviour of the system, leading to theoretical Evolutionary Model of Public Data Ecosystems (EMPDE) that depict the evolutionary trajectory of PDEs. Previous research [21] has proposed a typology of PDEs, their conceptual model with the elements that make up these ecosystems and identified six evolutionary generations, along with their constituent elements. Designed as a result of a systematic literature review on the topic spanning three decades, this model can be considered theoretically robust, but it requires empirical validation to enhance its practical applicability. Thus, **the aim of this study is to validate the theoretical Evolutionary Model of Public Data Ecosystems through a real-life examination in European countries.**

Specifically, our study validates the proposed generations and associated elements of PDE, including components and relationships, stakeholders, actors and their roles, data types, processes, and activities, and/or data lifecycle phases, with the aim to identify the formative characteristics of different evolutionary generations of PDE.

The rest of the paper is structured as follows: Section 2 provides the background of the study, including the theoretical model of the evolution of public data ecosystems, Section 3 presents the empirical research, including the research methodology, the model validation results and the revised model. Section 4 establishes a discussion, while Section 5 concludes the paper and identifies directions for future research.

## 2      Evolution of public data ecosystems

Public data ecosystems, as previously noted, are multidimensional [23-25] and dynamic [14-15], evolving over time both to different taxonomies, elements shaping them and relationships between these elements [21, 26-28].

Although this evolution is responsive to changes in technology, policy, and user needs [29], there has been little consensus in previous research regarding the factors or elements that contribute to the development / evolution of public and open data ecosystems. Some authors highlight the involvement of stakeholders beyond the public sector in the ecosystem and its development [20], while others focus on performance comparisons and best practices [13,17].

Heinz et al. [30] exploring the evolution of data ecosystems encompasses three main areas: (1) innovation, (2) engineering, and (3) collaboration. Others characterize this evolution by phases, where initially, they are designed around a new technology, which then is adopted by users from various sectors, which is compliant with trust-building stressed by Gelhaar and Otto [31], and finally, the ecosystem expands and grows [32]. To address this complexity, Lnenicka et al. [21] introduced a model delineating the evolutionary trajectory of PDEs, identifying six generations characterized by five meta-attributes / meta-characteristics: (1) formative components and relationships, (2) stakeholders, (3) actors and their roles, (4) data types, and (5) processes and activities, and associated data lifecycle phases (Fig. 1). This model, stemming from a systematic literature review, aims to provide clarity on the evolution of PDEs across various dimensions.

**The first generation**, called "**Raw Data-centred Generation**" (typically spanning from the 1990s to 1991/1993), focused on standalone data infrastructures primarily managed by governments, with a key emphasis on digitization. **The second generation**, called "**Geospatial Data-centred Generation**" (typically spanning from 1991/1993 to 2000s), witnessed the rise of Spatial Data Infrastructures (SDIs) and active engagement from citizens and academia to enhance innovation and data accessibility. Their involvement aimed to foster innovation, enhance data quality, and facilitate data visualization for improved accessibility to a broader spectrum of users. **The third generation**, referred to as the "**Public Data Sharing Generation**" (typically spanning from the 2000s to 2007/2009), intricately linked to the advancement of e-government, the proliferation of ICT, and the widespread adoption of the Internet among citizens and businesses, witnessed the formulation of initiatives for standardizing public data and encouraging data sharing and co-creation efforts.

| Characteristics forming the evolution of generations | Evolutionary generations of public data ecosystems | | | | | |
|---|---|---|---|---|---|---|
| | 1. GENERATION raw data-centred | 2. GENERATION (geo)spatial data-centred | 3. GENERATION public data sharing | 4. GENERATION open (government) data | 5. GENERATION public data-driven | 6. GENERATION intelligent public data |
| Components and relationships | Data resources, Data infrastructures | Data policy and other related policies, Data lifecycle management (data governance) | Technologies (platforms, tools, and services), Technical standards and guidelines | Data-related competencies and skills, Dynamics of processes and activities | Big and open linked data, cloud computing, External pressures – political, economic, environmental, ethical, and legal | Intelligent algorithms, machine learning, artificial intelligence, and natural language processing tools |
| Stakeholders | Governments (national), Public organizations | Governments (regional, local), Citizens, Academia | Private organizations, NGOs | Developers | International stakeholders | Networks, partnerships etc. |
| Actors and their roles | Data producers, Data owners, Internal data users | Data providers, External data users | Policies, laws, and rules parties, Data publishers | Ecosystem orchestrators, Data prosumers, Data intermediaries | Infrastructure providers, Platform providers, Service providers | Ecosystem managers, Data curators, Data consultants |
| Data types | Data (no specific type) | (Geo)spatial data | Public sector data, Metadata | Open data, Linked data | Big data, Real-time (stream) data | Intelligent (smart) data |
| Processes and activities, data lifecycle phases | Data generation, Data transfer | Innovation, Data quality, Data processing, Data visualization | Support, Data sharing, Participation, Collaboration | Transparency, Openness, Engagement, (Re)use | Data mashing, Impact, Decision-making, Sustainability | Resilience, Interoperability, Storytelling |

**Fig. 1.** A six-generation model of the evolution of public data ecosystems [21]

The first three generations were largely focused on data production and disclosure, but with limited interaction between data activities and stakeholders, so that there was virtually no open data ecosystem within their structures. Subsequent generations, in turn, have explored and incorporated elements of stakeholder and data interaction, giving rise to open data ecosystems.

**The fourth generation**, known as the "**Open (Government) Data Generation**" (typically spanning from 2007/2009 to 2013/2015), marked the inception of open data ecosystems, focusing on transparency, engagement, and the development of new applications and services. Users and developers played pivotal roles, being instrumental in shaping these ecosystems. The **fifth generation**, referred to as the "**Public Data-driven Generation**" (typically spanning from 2013/2015 to 2020s), introduced new components, including big data, Linked Open Data (LOD), and cloud computing to enhance decision-making processes and societal impact. Notably more open than its predecessors, this generation incorporated additional stakeholders such as international organizations, introducing external pressures that influence and shape the ongoing development of the ecosystem. **The sixth generation,** known as the "**Intelligent Public Data Generation**" (typically spanning from the 2020s to the present), integrates intelligent algorithms, AI, machine learning (ML), and natural language processing (NLP) tools into the landscape of PDEs, revolutionizing data processing and user interactions.

Intelligent (smart) public data becomes a cornerstone, fostering ecosystem resilience and interoperability. The advent of new actors and roles, such as ecosystem managers, data curators, and data consultants, contributes to more efficient management of PDEs. It's important to note that the sixth generation is forward-looking, derived from a literature review, where some studies, however, have taken a more conceptual approach. Therefore, we can expect most of the changes from empirical testing of this model to occur within this generation. We also assume that for many countries this generation may not currently exist.

## 3 Public data ecosystem evolution. Empirical research

### 3.1 Research and data collection methods

To empirically validate the six-generation model proposed by Lnenicka et al. [21], an expert assessment questionnaire with European national experts was conducted in sample countries. Although expert judgement as a research method has been criticised in literature due to validity concerns [33], this can be mitigated in the design of the expert survey by avoiding ambiguous terms to limit interpretations and by examining the internal consistency between experts and collected data [34].

To this end, we developed a questionnaire / protocol in which national experts were asked (1) to examine the **validity of the identified six generations** by determining their existence and relevance of each generation in his/her country, (2) the start and end time of each generation as well as its duration constituting a **temporal analysis**, (3) their opinion on the potential influence of regional or local-level "generations" and how citizens at different levels interact with these generations of PDEs, (4) the potential **additional country-specific generations** their country's PDE can be characterized by and, finally (5) the **existence and importance assessment of meta-characteristics** of these generations in the selected country's PDEs. This protocol is publicly available through Zenodo[1]. In addition, being interested in the next generation PDE, we also collected information on *what is expected to influence public data ecosystems in the future*. Response options in questions regarding the relevance of each generation or the meta-characteristics included in each generation being acceptability tasks used a 5-point Likert scale: "*yes – very important*", "*yes – important*", "*yes, but of little importance*", "*yes, but not very important*, "*no*". For a temporal analysis, where experts were asked to analyse the period when each generation began and ended, data sources recommended to perform this task included national documents, and respondents' personal knowledge.

Five countries were chosen to validate the proposed model based on their experience and achievements in the public data domain, as well as the influence of enabling technologies and concepts that have shaped evolution over the years. To select the most representative countries, we reviewed the relevant documents that provided countries assessments in terms of (1) data infrastructures and data-related services represented in e-government, digital, and ICT development indices, such as the *E-Government Development Index*, *ICT Development Index*, *Network Readiness Index*, and *EU's*

---
[1] https://doi.org/10.5281/zenodo.11146461

*eGovernment Benchmark*; (2) open (government) data, and indices of data availability and impacts, such as the *Open, Useful and Re-usable data Index (OURdata)*, *Open Data Maturity Index*, and *Open Data Inventory*; and (3) technologies and approaches shaping work with public data, such as the *Global Cloud Ecosystem Index*, *Global Cybersecurity Index*, *AI Index*, and *Government AI Readiness Index*. To ensure diversity, our selection encompassed both highly competitive and less competitive countries. This approach aimed to prevent an exclusive focus on the most successful nations, promoting a nuanced and inclusive comprehension of varied practices. Simultaneously, avoiding only the lowest-performing countries enabled us to consider and integrate best practices into the study. This selection process was complemented by convenience sampling, where the availability of experts determined our sample. Consequently, the five countries comprising our sample are Latvia, Serbia, Spain, Czech Republic, and Poland.

The term "expert" in the context of this study is defined in line with [16] and refers to an individual possessing profound expertise encompassing both knowledge of the subject matter and an understanding of the contextual nuances specific to a particular country, which could influence the outcomes. This expertise may stem from either the expert's country of origin or the country where they are actively engaged or employed in open data initiatives. Crucially, the individual must hold a minimum of a master's degree in a field relevant to PDEs and/ or (impact of) emerging technologies in the public sector and demonstrate familiarity with diverse domains such as business and management, political sciences, law, computer sciences, among others. Additionally, the expert should boast a minimum of five years of comprehensive research and practical experience pertaining to data projects and/or initiatives within the public administration of the relevant country. Moreover, in our study, selected national experts were those who participated in the creation and design of the Evolutionary Model of Public Data Ecosystems (EMPDE). Thus, all concepts and characteristics to be measured were clear to all experts who had sufficient abilities, skills, and knowledge to answer the assessment questionnaire. In addition, the experts' responses were validated with information provided in the official documents of the selected countries. These methods ensure the reliability of experts' response to the expert assessment questionnaire.

### 3.2 Analysis of results

**Existence and importance assessment of six generations of public data ecosystems and their characteristics in selected countries.** The results show that all six PDE generations are present across the sample countries, with a few exceptions for individual countries (Table 1). Specifically, experts noted the absence of the initial (1st and 2nd) and the most advanced (5th and 6th) generations in Serbia and Spain, where the latter - the absence of the 6th generation aligns with our earlier expectations.

Table 1. An existence and importance assessment of six generations in 5 countries

| Generation / existence & importance | Yes – very important | Yes – important | Yes – little important | Yes – not important | Does not exist |
|---|---|---|---|---|---|
| 1st generation | 0 | 1 of 5 | 1 of 5 | 2 of 5 | 1 of 5 |
| 2nd generation | 2 of 5 | 2 of 5 | 0 | 0 | 1 of 5 |
| 3rd generation | 3 of 5 | 2 of 5 | 0 | 0 | 0 |
| 4th generation | 4 of 5 | 1 of 5 | 0 | 0 | 0 |
| 5th generation | 1 of 5 | 3 of 5 | 0 | 0 | 1 of 5 |
| 6th generation | 1 of 5 | 1 of 5 | 2 of 5 | 0 | 1 of 5 |

The level of importance attributed to these generations, however, varies both across countries and generations. The 4th generation was found the most important, where 4 of five countries found it to be very important and one country - important, followed by the 3rd generation and the 2nd generation. The 5th generation was predominantly considered important, with one country emphasizing it as very important, while in another, it was found as non-existent (Spain).

By analysing the evolutionary trajectory of PDE generations, we found that the start and end dates and durations of these generations (Fig. 2) vary slightly from country to country. They also differ from the dates reported in the literature, which is somewhat what we have expected and considered one of the main reasons for empirical validation (Section 2).

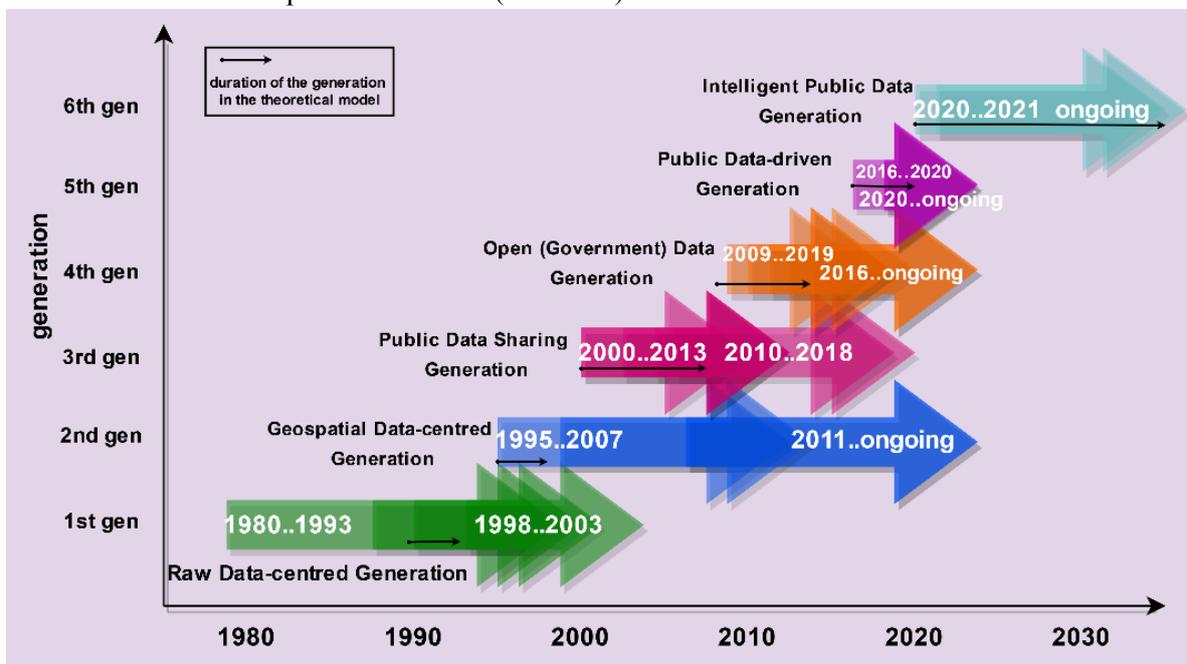

**Fig. 2.** Generational periods of public data ecosystems

Thus, in some countries, the first generation started 10 years earlier, that is, in the 1980s, and continued until 1998 and 2003 depending on the country. For other generations, the start dates generally coincided with those determined based on the literature review, however, the end dates turned out to be significantly different, especially for generations one through five. Probably the most interesting in this context is the 2[nd] generation, since, although it existed in all analysed countries until 2011 as an independent generation, it continues to exist today, albeit in a different form. Today, it predominantly exists as a specific subtype of PDEs, such as a local government open data ecosystem or a smart city data ecosystem.

Our examination of characteristics outlined in the theoretical model (Fig. 1), confirms their significance in driving the evolution of PDE generations (Table 2). Most of the characteristics are assessed as "*important*" and "*very important*" across all PDE generations (except for the 1[st] generation), emerging as recurrent themes across successive generations, albeit with some nuances observed in the initial iteration. This indicates their relevance as fundamental pillars of PDE, regardless of slight deviations within certain PDE generations.

**Table 2.** An importance assessment of six generations in selected countries

| Characteristics shaping generation | 1st gen. | 2nd gen. | 3rd gen. | 4th gen. | 5th gen. | 6th gen. |
|---|---|---|---|---|---|---|
| **Components & relationships** | *i/l* | *v/i* | *v/i* | *v/i* | *i/v* | *i* |
| **Stakeholders** | *l/n* | *i/l* | *i/v* | *vi* | *v/i* | *v/i* |
| **Actors & their roles** | *l/n* | *i/v* | *v/i* | *i/v* | *v/i* | *v/i* |
| **Data Types** | *n/l* | *i/v* | *v/i* | *i/v* | *i/v* | *i/v* |
| **Processes & activities, data lifecycle** | *l/i* | *i/vi* | *i/vi* | *v/i* | *i/v* | *i/v* |

*v* - very important, *i* – important, *l* - low importance, *n* – not important
red – rather not important, yellow – rather little important, blue – rather important, green – rather very important

For the first generation, **components and relationships** and **processes and activities and data lifecycle** were the most important meta-attributes, with **data types**, **actors and their roles,** and **stakeholders** found to be of the least importance. Subsequent generations witnessed a shifting landscape wherein all meta-attributes became important, aligning with the evolving focus of each era. Notably, the importance attributed to dynamics such as **stakeholders' roles** undergoes a discernible progression, with **stakeholders** ascending to the greatest importance by the 4[th] generation. Conversely, the significance of **data types** within the PDE framework experiences a notable transition, with their pivotal role emerging prominently from the 2[nd] generation onwards.

**Revised six-generation model of the evolution of public data ecosystems.** The qualitative analysis of the evolution of the different characteristics is shown in Fig. 3.

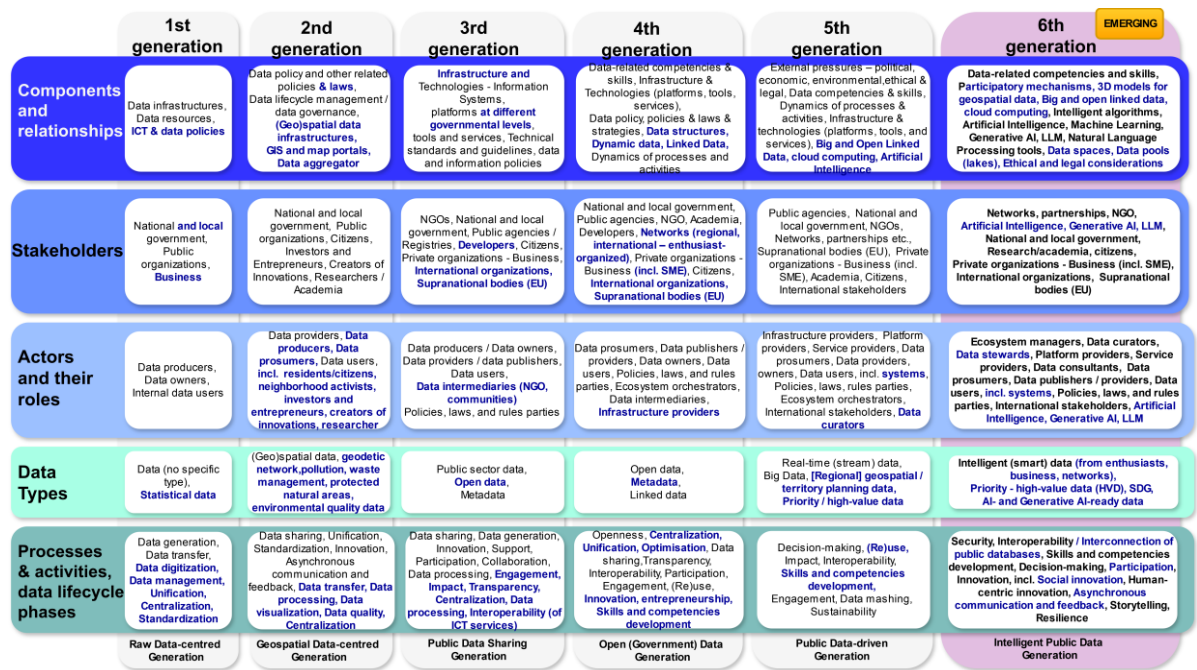

**Fig. 2.** A validated six-generation model of the evolution of public data ecosystems (blue is used for new elements; for better resolution see https://doi.org/10.5281/zenodo.11146461).

According to the expert panel used in this research, the 1st PDE generation was driven by the need to invest in **data infrastructure** and **resources** and **ICT policies**, primarily driven by **national and local governments**, **public organizations**, and **businesses**. The key actors include **data producers**, **owners**, and **internal data users**, focusing on **managing** and **standardizing data processes.** While the theoretical model does not specify the data types handled by PDEs of this generation, our analysis identifies **statistical data** as the predominant category for this generation, managed through **digitization**, **transfer**, **unification**, and **standardization** efforts to ensure effective utilization across sectors and stakeholders.

The 2nd generation centres around (geo)spatial data infrastructures and GIS portals associated with the regulatory **framework** (**data policy**, **lifecycle management**, and **governance**) catering to **citizens**, and **researchers**, **national and local governments**, **public organizations**, who later were accompanied by **investors and entrepreneurs**, **creators of innovations**, and **neighbourhood activists**. Actors encompass **data providers**, **producers**, **prosumers**, and **external users**, including **residents**, **activists**, and **academia**, managing diverse **geospatial data** collections that later were expanded to **geodetic networks** and **environmental quality data**. Processes, in turn, included **data sharing**, **processing**, **transfer**, **visualization**, **innovation**, and **quality assurance**, facilitated by **asynchronous communication channels for feedback**. The focus lies on optimizing **data utilization** and **governance** across sectors while fostering **innovation** and **collaboration**, later expanded to **participation mechanisms**, including in the city's development processes.

The 3rd generation is characterized by a robust **infrastructure** and **technological framework,** accompanied by **platforms at different governmental levels, tools and services** governed by **technical standards** and **data policies** across governmental levels. Stakeholders include **NGOs**, **developers**, **citizens**, and **international organizations** and **supranational bodies** (e.g., EU). Actors within this generation who play distinct roles are **data intermediaries** that complement the list of more traditional actors observed during previous generations. **Policies and regulations** drive ecosystem dynamics, while data types encompass **public sector and open data** managed with **metadata standards**. Processes focus on **participation**, **collaboration**, **engagement**, **transparency**, **impact creation** and **interoperability** in ICT services to optimize data utilization.

The 4th generation saw the creation of **networks** - regional and international, **ecosystem orchestrators** and **infrastructure providers** as new stakeholders and actors, respectively, improving their data structures and introducing **dynamic** and **linked data** fostering **openness**, **centralization**, **unification**, **interoperability**, **sharing**, **transparency**, **participation,** and **engagement**. Processes, in turn, aimed at **openness** and implying **(re)use**, **data-related competencies and skill development**, **entrepreneurship**, and **optimization**, emphasizing the importance of **dynamic processes**, including **data management** and increased attention towards data structures.

The 5th generation is characterized by the early integration of **emerging technologies** (Big and Linked Open Data, cloud computing and AI) into the PDE system, and the introduction of **data curators** as new actors for organising and integrating data coming from different sources aimed at putting them available for their reuse and conservation. Although the list of actors has expanded to include **platform providers**, **service providers**, **international stakeholders** in addition to data curators, no new stakeholders have been observed in this generation. However, it observed **external pressures** from political, economic, environmental, ethical, and legal domains, where the latter is associated with the increase of data volumes, including **Big Data**, **real-time data (stream data)**, and **priority/high-value data** that became central data types for this generation, while processes focused on **decision-making**, **data mashing** and **sustainability** that complemented those that saw the rise within previous generations, namely impact assessment, interoperability, (re)use, skills development, and engagement.

Finally, the 6th generation introduces the focus on **participatory mechanisms**, and **advanced technologies** such as 3D models for geospatial data, which expects to witness integration of intelligent algorithms including AI, Machine Learning, and NLP tools, Large Language Models and Generative AI. Moreover, **Artificial Intelligence**, **Generative AI**, and **LLM** are also becoming actors and stakeholders, complemented by actors such as **ecosystem managers**, **data consultants** and **data stewards**. The latter - data stewards - become essential to ensuring the quality and fitness for the purpose of an organization's data assets, including the metadata for those data assets. Additionally, some countries observe the integration of the concept of **data spaces and pools (data lakes)**, facilitating efficient data management and utilization. Data types encompass **intelligent (smart) data**, **priority/high-value data**, **SDG-related data**, and **AI- and Generative AI-ready data**, catering to diverse needs and applications. The processes focus on **security**, **storytelling**, and **resilience,** while **skills development**, **decision-making**, **participation in urban development processes**, fostering **human-centric and** **social innovation,** and **interoperability** remain crucial within this generation. The latter - interoperability, especially the interconnection of public databases, effective bi-directional communication, innovation, and resilience emerge as central elements. This generation emphasizes leveraging cutting-edge technologies and participatory approaches to drive innovation and resilience in data ecosystems.

In short, according to the panel, while the most important characteristics in the first three generations of PDE can be seen as aimed at creating an open data ecosystem (and OGD portals as a central entry point), starting with the 4$^{th}$ generation of PDE, the most important characteristics of PDE are focused on improving systems created not only through the innovation brought about by new technologies, but also through the need to establish requirements and rules to solve their problems.

## 4   Discussion

The first generation underscores a lack of documentation and information regarding the undertaken initiatives, alongside a lack of strategic approach at the system level. This indicates minimal interest from the research community in digitization activities within the public sector at that time, resulting in challenges in reconstructing knowledge from this era, leading to a scientific blind spot.

Generations two through five are characterized by progressive centralization efforts. This is evidenced by the integration of functionalities into dedicated platforms, commonly known as one-stop-shops, primarily situated at the central government level. For instance, in Latvia's 2$^{nd}$ generation PDE, a single access point for geospatial information and services was established, gradually evolving into a more centralized approach in subsequent phases due to the fact that the national open data portal is more compliant with the current needs, trends, and expectations of both publishers and users. Similar centralization initiatives were observed in Poland, where data and information provision were consolidated at the government level, exemplified by the transition to the central portal *gov.pl*, eliminating separate websites and portals of individual public institutions (following a similar approach as the UK did a decade ago).

The 6$^{th}$ generation of PDEs is currently at an early stage of development, with limited exploration and initial regulatory developments in government sectors worldwide. This generation is characterized by forward-looking discussions, with emerging technologies anticipated to play pivotal roles in shaping its evolution. Although these emerging technologies vary in number and nature, one of the most widely discussed and adopted technologies is AI. This is also supported by the fact that many countries, including those included in the sample, are launching development strategies and plans tailored to this technology. For example, in February 2023, the Serbian government adopted *Ethical guidelines for the development, application, and use of reliable and responsible AI*. The Czech Republic, Latvia, and Poland have benefited from investments supported by the EU's *Recovery and Resilience Facility*, which funds projects focused on digital transformation and policies for the next generation. Spain has launched an *AI sandbox* to help start-ups adapt to the EU's *AI Act*. Thus, in the following section, we will focus primarily on AI and discuss the potential impact of AI on the 6$^{th}$ PDE generation, leveraging insights from our study.

## 4.1 Emerging technologies for a new generation of public data ecosystems

In the context of the 6$^{th}$ generation, there's a continued emphasis on **data-related competencies, skills**, and **dynamic processes**, alongside components related to **policies**, **infrastructure**, and **data structures**, including **data lakes** and **data spaces**. This generation encompasses (or is expected to encompass) **intelligent (smart) data** sourced from various stakeholders beyond **public agencies**, including **enthusiasts**, **businesses**, and **networks**, and **priority data** categories that range from **high-value datasets** to **Sustainable Development Goals (SDG).** However, the foremost focus is on emerging technologies such as **intelligent algorithms**, **NLP tools**, **AI**, and **Machine Learning**, which necessitate profound **ethical and legal considerations**.

**AI** and **Generative AI**, in turn, are poised to play pivotal roles in both **automation** and **augmentation**. These advancements encompass a spectrum of functionalities, including **processing tools**, **recommendation tools**, and **analysis tools** for automation, as well as **individualized**, **enriching**, and **ideation feedback mechanisms** for augmentation, spanning **individual**, **peer-to-peer**, and **collective levels** of interaction. This also aligns with [35], the discourse outlined in [2] emphasizing the need for the AI-driven transformation, [36] on the "AI-enhanced online ideation", as well as some evidence of chatbots integrated into open data ecosystem [37-38].

**LLMs** and **generative AI** can empower **data exploration** and **visualization tools**, facilitating intuitive **interaction** with datasets, thereby contributing to better data understanding even for novices - users with limited data literacy levels. These tools can have the potential to **generate descriptions in natural language**, **visualizations**, or even **interactive simulations**, enhancing understanding and exploration of complex datasets and their transformation into tangible results.

Since LLMs and generative AI can generate rich, coherent text content based on input prompts or seed data, we can expect that this capability will be leveraged to create **summaries**, **annotations**, **explanations**, or **additional context** for open datasets, enhancing their **interpretability** and **usability** for various stakeholders. At the same time, this can be used at earlier stages to better **document** datasets, extracting data and information from them before they are published, i.e., **tags**, **descriptions,** etc. by augmenting data publishers' input and thereby improving both data **findability**, **understandability** at earlier stages with further contribution to better **data linkage** through the common characteristics extracted from these datasets.

Consequently, we can expect AI, generative AI and LLM to become integral part of the public data ecosystem of the sixth generation. For example, in Poland, the *Policy for the Development of Artificial Intelligence in Poland from 2020* includes provisions for the further development of open data systems and a focus on trustworthy AI.

## 4.2 New technologies, old problems, and some paradoxes

While AI and Generative AI offers vast potential across numerous domains, its effective utilization hinges on the readiness of (open) data for the capabilities of generative AI technologies. This necessitates a paradigm shift, introducing new prerequisites to ensure data quality, security, and ethical considerations.

To this end, several critical considerations must be addressed to make open data AI- and Generative AI-ready, which in many cases are also valid for previous generations, where the level of importance has increased within the 6$^{th}$ generation. Since the AI and Generative AI models rely on the data they are trained on, **data quality** is paramount, encompassing not only **completeness**, but also other data quality dimensions, including **accuracy**, **reliability,** and comprehensive coverage of relevant information, including **metadata**. This involves data **preparation** and **pre-processing**, involving **cleaning**, **normalization**, and **standardization** to ensure compatibility with (generative) AI technologies. **Metadata** and associated documentation serve as indispensable sources for understanding data context, structure, and limitations, underscoring their pivotal role in facilitating effective data interpretation and utilization. Paradoxically, while AI and Generative AI may contribute to enhancing data quality and metadata, they inherently rely on their pre-existence. Thus, data must meet these criteria independently to optimize their utilization. Moreover, the more qualitative are the above elements - data quality, metadata etc., the better the effect of AI and Gen AI from application even to strengthen them.

Given AI and generative AI potential to generate synthetic data that could inadvertently reveal private information if not properly handled, **privacy and ethical considerations** assume heightened significance with the emergence of these technologies. Therefore, stringent measures, including **anonymization** and measures against de-identification, are imperative to mitigate privacy risks effectively.

In the context of PDEs, where data is made available to the public and used for various purposes including AI-driven decision-making, the transparency and interpretability of AI systems become crucial. **Explainable AI (XAI)** is important for ensuring accountability, understanding the reasoning behind AI decisions, and building trust among users and stakeholders. In recent years, there has been a shift in focus towards the predictive power of AI algorithms, often neglecting the need for understanding the underlying mechanisms and interpretations of AI decisions. However, within public and open data ecosystems, where ethical considerations are paramount, the revival of XAI is necessary to address these challenges. Integrating XAI principles into AI services within next-generation data ecosystems can contribute to the enhancement of their transparency and interpretability. This, in turn, **reduces potential ethical and legal issues** associated with AI-driven decision-making. By ensuring that AI systems are understandable and explainable, users can mitigate risks and improve the acceptability and trustworthiness of AI-based solutions among users and stakeholders.

As such, **community engagement** and **collaboration** play a pivotal role in addressing these challenges. By fostering dialogue and collaboration among users, developers, and stakeholders, we can identify potential issues and enhance the quality and diversity of open data. This collaborative approach not only enriches the data ecosystem but also broadens the utility and applicability of AI and generative AI technologies.

## 5    Conclusions

The study validates a theoretical Evolutionary Model of Public Data Ecosystems (EMPDE) [2] through real-life examination in five European countries, providing insights into the dynamics and variations of these ecosystems across different meta-attributes: (1) formative components and relationships, (2) stakeholders, (3) actors and their roles, (4) data types, and (5) processes and activities, and associated data lifecycle phases. Our empirical research identifies the relevance of all generations of public data ecosystems across selected European countries (with some exceptions for individual countries) and confirms the characteristics forming these evolution models, as identified in the theoretical model stemming from a literature review. However, we find that the list of characteristics and elements shaping different generations of PDEs was not exhaustive. In this study, we complement this list with elements identified in real-world public data ecosystems. Similarly, by analysing the evolutionary trajectory of PDE generations, we found that the lifespan of most generations as reported in the literature differs from actual dates and lifespans of generations, which was true for all generations except the sixth, which is what we expected and considered one of the main reasons for empirical testing. Thus, by empirically validating the theoretical evolutionary model of public data ecosystems through a real-world study in European countries, we have improved its accuracy, making it compliant with the real-world environment.

Among generations, we focused on the $6^{th}$ generation of PDEs, which represents a paradigm shift fuelled by emerging technologies such as cloud computing, AI, NLP, Generative AI, and LLM. These technologies have the potential to contribute to both automation and augmentation of business processes within these ecosystems. We therefore briefly explored the potentialities within the sixth forward-looking generation and their implications for effective PDE systems, providing avenues for future research. The study highlights the potential impact of emerging technologies, such as AI, generative AI, and LLMs that are poised to become indispensable components of PDE, assuming roles as both actors and stakeholders. Their active engagement and advocacy promise to shape the future trajectory of the ecosystem, driving innovation, promoting transparency, and fostering inclusivity. By consuming open data, AI and Generative AI contribute to the generation of insights, the development of applications, and the advancement of knowledge. Simultaneously, they advocate for data quality, reliability, and fairness, thereby safeguarding the integrity of the ecosystem as a whole.

However, these new technologies also pose new challenges that need to be addressed. The first concerns data quality, including completeness, accuracy, reliability, and comprehensive coverage of relevant information and metadata, as AI relies on their pre-existence. The second concerns data privacy and security, including ethical considerations, where integrating XAI principles into AI services can increase transparency and interpretability, reducing ethical and legal issues associated with AI-driven decision-making.

However, as our findings suggest, AI is not the only emerging technology that is expected to shape future PDEs, so in our future research we will expand the scope covering other technologies and associated topics that span from cloud computing to Big Data and data lakes. In addition, our future research will explore which AI solutions and approaches are suitable for different stakeholders and their roles in the public data ecosystems, focusing on specific activities and business processes that correspond to these roles. Overall, this topic deserves deeper analysis to develop effective PDE to build trust in data and collaborative governance models, and improve data use, and stakeholder participation, including in government decision-making.

## ACKNOWLEDGMENTS

We acknowledge the use of ChatGPT-3.5 to (partly) word selected parts of this paper.

15. Wilson, L. (2019). Understanding the data ecosystem. In *Data-driven Marketing Content* (pp. 11-21). Emerald Publishing Limited, Bingley.
16. Lnenicka, M., Nikiforova, A., Clarinval, A., Luterek, M., Rudmark, D., Neumaier, S., Kevic, K. & Bolívar, M. P. R. (2024). Sustainable open data ecosystems in smart cities: A platform theory-based analysis of 19 European cities. *Cities*, 148, 104851.
17. Styrin, E., Luna-Reyes, L. F., & Harrison, T.M. (2017). Open data ecosystems: An international comparison. *Transforming Government: People, Process and Policy*, 11(1), 132-156.
18. Hendriks, P. H., Dessers, E., & Van Hootegem, G. (2012). Reconsidering the definition of a spatial data infrastructure. *International Journal of Geographical Information Science*, 26(8), 1479-1494.
19. Mulder, A. E., Wiersma, G., & Van Loenen, B. (2020). Status of national open spatial data infrastructures: A comparison across continents. *International Journal of Spatial Data Infrastructures Research*, 15, 56-87.
20. Vancauwenberghe, G., & van Loenen, B. (2018). Exploring the emergence of open spatial data infrastructures: analysis of recent developments and trends in Europe. In *User Centric E-Government. Integrated Series in Information Systems* (pp. 23-45). Springer, Cham.
21. Lnenicka, M., Nikiforova, A., Luterek, M., Milic, P., Rudmark, D., Neumaier, S., Kević, K., Zuiderwijk, A., Rodríguez Bolívar, M. P., Understanding the development of public data ecosystems: from a conceptual model to a six-generation model of the evolution of public data ecosystems (2023), *SSRN*, http://dx.doi.org/10.2139/ssrn.4831881
22. Aaen, J., Nielsen, J. A., & Carugati, A. (2022). The dark side of data ecosystems: A longitudinal study of the DAMD project. *European Journal of Information Systems*, 31(3), 288-312.
23. Beverungen, D., Hess, T., Köster, A., & Lehrer, C. (2022). From private digital platforms to public data spaces: Implications for the digital transformation. *Electronic Markets*, 32(2), 493-501.
24. Hrustek, L., Mekovec, R., & Alexopulos, C. (2023). Concept for an open data ecosystem to build a powerful data environment. In *Information Systems: 19th European, Mediterranean, and Middle Eastern Conference, EMCIS 2022* (pp. 251-263). Cham, Springer Nature Switzerland.
25. Ruijer, E., Dingelstad, J., & Meijer, A. (2023). Studying complex systems through design interventions probing open government data ecosystems in the Netherlands. *Public Management Review*, 25(1), 129-149.
26. Guggenberger, T. M., Möller, F., Haarhaus, T., Gür, I., & Otto, B. (2020). Ecosystem Types in Information Systems. In *Proceedings of the 28th European Conference on Information Systems (ECIS)* (p. 45).
27. Lis, D., & Otto, B. (2021). Towards a taxonomy of ecosystem data governance. In *Proceedings of the 54th Hawaii International Conference on System Sciences* (pp. 6067-6076). ACM.
28. Gelhaar, J., Groß, T., & Otto, B. (2021). A taxonomy for data ecosystems. In *Proceedings of the 54th Hawaii International Conference on System Sciences* (pp. 6113-6122). ACM.
29. Jetzek, T. (2017). Innovation in the open data ecosystem: Exploring the role of real options thinking and multi-sided platforms for sustainable value generation through open data. In *Analytics, Innovation, and Excellence-Driven Enterprise Sustainability* (pp. 137-168). Palgrave Macmillan, New York.
30. Heinz, D., Benz, C., Fassnacht, M., & Satzger, G. (2022). Past, present and future of data ecosystems research: A systematic literature review. In *PACIS 2022 Proceedings* (p. 46). AIS.
31. Gelhaar, J., & Otto, B. (2020). Challenges in the emergence of data ecosystems. In *PACIS 2020 Proceedings* (p. 175). AIS.
32. Otto, B., & Jarke, M. (2019). Designing a multi-sided data platform: findings from the International Data Spaces case. *Electronic Markets*, 29(4), 561-580.
33. Khatri, N., & Ng, H. A. (2000). The role of intuition in strategic decision making. *Human relations*, 53(1), 57-86.
16